\documentclass[11pt]{article}
 \pdfoutput=1 
 \usepackage{geometry}                
 \usepackage{graphicx}
 \usepackage{epstopdf}
 \usepackage{amssymb}

\usepackage{amsmath}
\usepackage{bbm}
\usepackage{pifont}
\usepackage{amsfonts}
\usepackage{hyperref}
\usepackage{mathrsfs}
\usepackage{pstricks}
\usepackage{color}
\usepackage{multicol}
\usepackage{cite}

 \title{\bf Instanton Fermionic Zero Modes of Heterotic\\ CP(1) Sigma Model}
 \author{Jin Chen\\ 
 \\ 
 {\em School of Physics and Astronomy, University of Minnesota}\\
 {\em 116 Church Street S.E. Minneapolis, MN 55455, USA}\\
 \\
 {\tt jinchen@physics.umn.edu}\\}

 \date{}                                           
 
\begin{document}
 \maketitle

\begin{abstract}
We discuss instanton fermionic zero modes in the heterotic $\mathcal{N}=(0,2)$ modification of the {CP(1)} sigma model in two dimensions. By calculating its chiral anomaly we prove that the number of fermionic zero modes is same as in the standard $\mathcal{N}=(2,2)$ {CP(1)} case, and determine their explicit form.
\end{abstract}
\section{Introduction}

It has been studied years ago that the supersymmetric generalizations of the
bosonic {CP(N-1)} sigma model is automatically an $\mathcal {N}\!=\!(2,2)$
supersymmetric theory by the virtue of that the target space
{CP(N-1)} is a K\"{a}hler manifold \cite{4}. 
Recently Edalati and Tong suggested and built an
$\mathcal{N}\!=\!(0,2)$ {CP(N-1)} heterotic sigma model by study of the low-energy
dynamics of vortex strings in $\mathcal{N}\!=\!1$ four-dimensional gauge
theories\cite{5}. Later Shifman and Yung formulated a geometric
representation for this heterotic model with the {CP(N-1)} target
space for bosonic fields and an extra right-handed fermion which
couples to the fermion fields of the $\mathcal{N}\!=\!(2,2)$ CP(N-1)
model\cite{1}, and thus breaks the supersymmetry to $\mathcal{N}\!=\!(0,2)$.

In this paper, we follow the geometric formulation to discuss the
fermionic zero modes of $\mathcal{N}\!=\!(0,2)$ heterotic CP(1) sigma model in the instanton background.  
Nontrivial homotopy group structure of target space CP(1) allows for 
the bosonic instanton background of the form  of a holomorphic function $\phi=\rho/(z-z_{0})$, 
where $\rho$ and $z_{0}$ are instanton size and center respectively.  
Under such background, the fermionic zero modes can be obtained 
by acting supercharges and superconformal charges to the instanton solution. Since the
heterotic  deformation of the standard {CP(1)} sigma model
only preserves half of four original supercharges, the supersymmetries and superconformal 
symmetries are partly broken. The remaining fermionic generators are not sufficient to generate
all fermionic zero modes from the instanton background.
Therefore we will find zero modes by solving Dirac equations. In
Section 2, we will calculate the chiral anomaly at first to give
the number of zero modes. In Section 3 we will explicitly solve
Dirac equations of zero modes and use the result of Section 2 to
pick up those linear independent modes. In the last Section we
generalize this result to {CP(N-1)} heterotic sigma models.\\

\section{Chiral Anomaly}
We start from the chiral anomaly of the standard CP(1) sigma model.
The notations are in accordance with the reference \cite{1}. The
Lagrangian of CP(1) is given as follows:
\begin{eqnarray}
\mathcal{L}&=&\frac{2}{g_{0}^2\chi^2}[\partial_{\mu}\phi^{\dagger}\partial^{\mu}\phi+i\bar{\psi}\gamma^{\mu}(\partial_{\mu}\psi-\frac{2}{\chi}\phi^{\dagger}\partial_{\mu}\phi\psi)+\frac{1}{\chi^2}(\bar{\psi}\psi)^2]\nonumber\\
&=&\frac{2}{g_{0}^2\chi^2}\partial_{\mu}\phi^{\dagger}\partial^{\mu}\phi+\frac{2i}{g_{0}^2}\bar{\xi}\gamma^{\mu}(\partial_{\mu}\xi-iA_{\mu}\xi)+\frac{2}{g_{0}^2}(\bar{\xi}\xi)^2
\end{eqnarray}
where $\psi=(\psi_R, \psi_L)^T$, $\xi\equiv\psi/\chi$ and $A_{\mu}\equiv
(i/\chi)\phi^{\dagger}\overleftrightarrow{\partial_{\mu}}\phi$ for
convenience to calculate the chiral anomaly. It has a chiral symmetry
and thus an axial current:
\begin{eqnarray}
j^{\mu
}_5=\frac{2}{g_{0}^2\chi^2}\bar{\psi}\gamma^{\mu}\gamma_{5}\psi=\frac{2}{g_{0}^2}\bar{\xi}\gamma^{\mu}\gamma_{5}\xi
\end{eqnarray}
The chiral anomaly is well-known\cite{3} as:
\begin{eqnarray}
\partial_{\mu}j^{\mu
}_5=\frac{1}{\pi}\epsilon^{\mu\nu}\partial_{\mu}A_{\nu}=\frac{2i}{\pi}\epsilon^{\mu\nu}\frac{\partial_{\mu}\phi^{\dagger}\partial_{\nu}\phi}{\chi^2}\,.
\end{eqnarray}
This anomaly term corresponds to the tadpole diagram:\\
\hspace*{5.5cm}\includegraphics[scale=1]{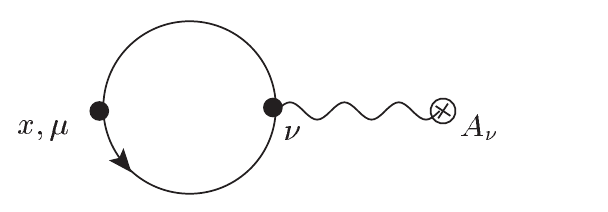}

Now we are going to consider the chiral anomaly of heterotic {CP(1)} sigma model. The Lagrangian of the heterotic model for bilinear fermion terms is deformed from Eq.\,(1) by adding terms with an extra right-handed fermion field $\zeta_{R}$ \cite{1}. Since only bilinear terms of fermions need to be considered when calculating one loop diagrams, we just write them up:
\begin{eqnarray}
\mathcal {L}_{biferm.}&=&\zeta_{R}^{\dagger}i\partial_{L}\zeta_{R}+[\gamma
g_{0}^{2}\zeta_{R}G(i\partial_{L}\phi^{\dagger})\psi_{R}+h.c.]
+\frac{2i}{g_{0}^2\chi^2}\bar{\psi}\gamma^{\mu}(\partial_{\mu}\psi-\frac{2}{\chi}\phi^{\dagger}\partial_{\mu}\phi\psi)\nonumber\\
&=&i\bar{\zeta}\gamma^{\mu}\frac{1+\gamma_5}{2}\partial_{\mu}\zeta+(\bar{\zeta}\gamma^{\mu}\frac{1+\gamma_5}{2}B_{\mu}\xi+h.c.)+
\frac{2i}{g_{0}^2}\bar{\xi}\gamma^{\mu}(\partial_{\mu}\xi-iA_{\mu}\xi)
\end{eqnarray}
where we define $\zeta=(\zeta^{\dagger}_R, 0)^T=(1+\gamma_{5})\zeta/2$ and $B_{\mu}=2i\gamma\partial_{\mu}\phi^{\dagger}/\chi^2$. It is easy to see $\gamma_{5}\zeta=\zeta$ and the Lagrangian is invariant under $\psi\rightarrow e^{i\alpha\gamma_5}\psi$ and $\zeta\rightarrow e^{i\alpha\gamma_5}\zeta$. Therefore the corresponding axial current is:
\begin{eqnarray}
j^{\mu}_5=\frac{2}{g_{0}^2}\bar{\xi}\gamma^{\mu}\gamma_{5}\xi+\bar{\zeta}\gamma^{\mu}\frac{1+\gamma_5}{2}\zeta
\end{eqnarray}

We will show that the chiral anomaly of the heterotic model will not be corrected by the extra $\zeta$ terms, and thus prove that the number of zero modes is the same as the standard {CP(1)} model. Actually, to calculate this anomaly one needs to consider diagrams as follows:\\
\includegraphics[scale=1]{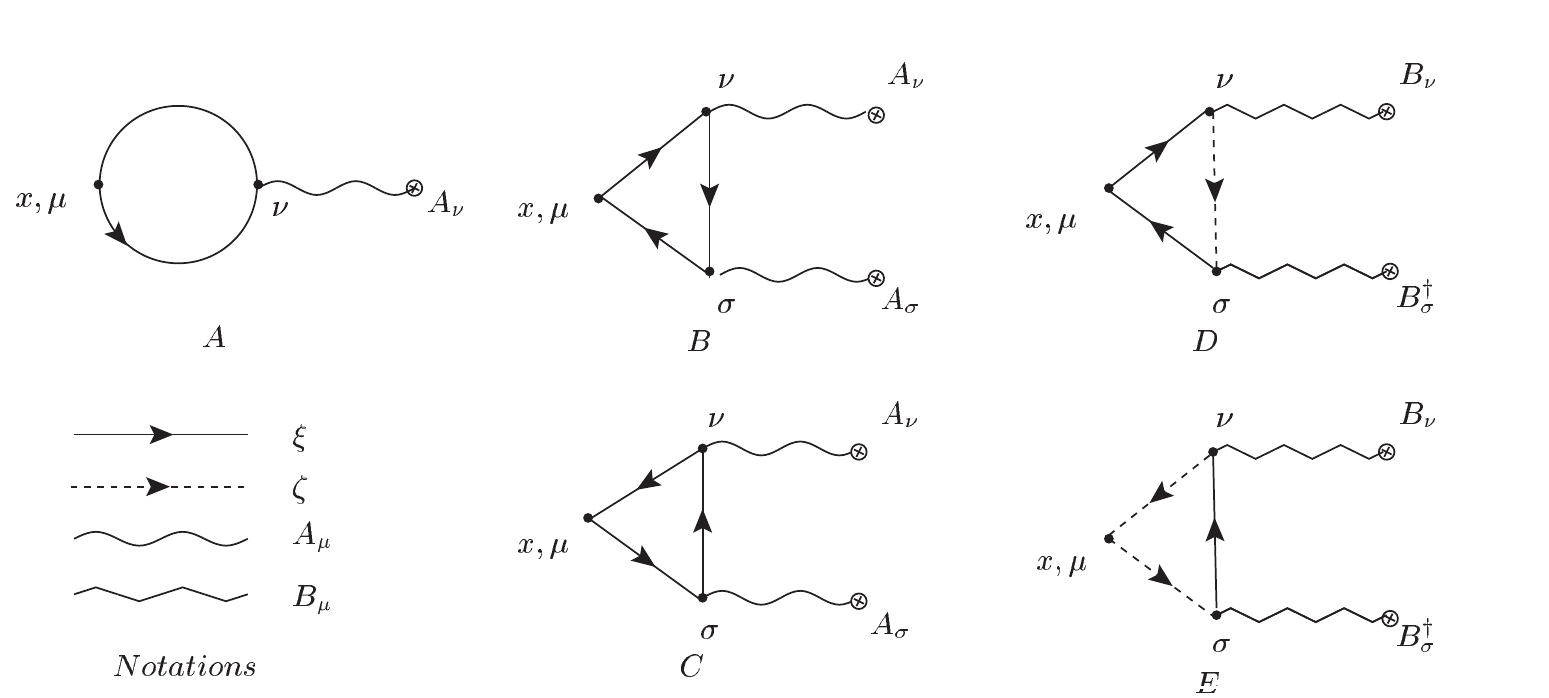}\\
The diagrams B and C cancel with each other because both of them are finite, and thus free to shift integral variables. The diagrams D and E also cancel each other. It can be proved by using momentum $k^{\mu}$ times the sum of these two diagrams, $i.e.$\\
\includegraphics[scale=1]{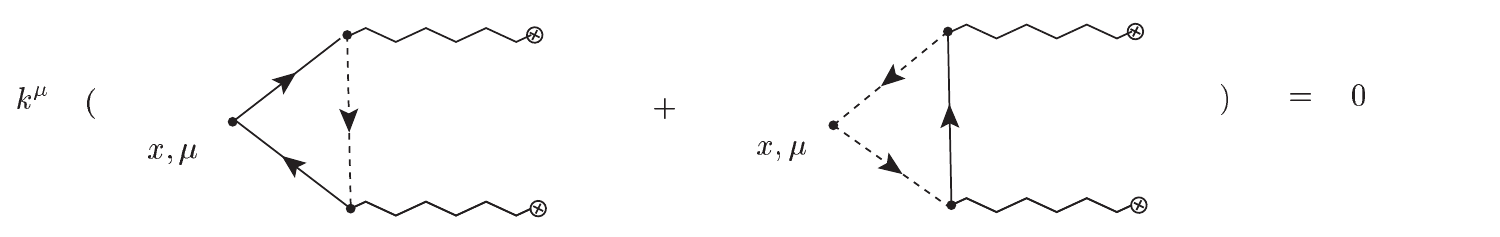}\\
By multiplying $k^\mu$, the original converged digrams D and E become two divergent parts respectively. However they are all logarithmically divergent, and thus still free to shift integral variables to prove the cancellation. Therefore the anomaly of heterotic case is still only from the tadpole diagram, and is the same as in the standard CP(1) case, see equation (3).\\

\section{Fermionic Zero Modes}

For the standard CP(1) model, $\psi$ has four zero modes, under the instanton background $\phi=\rho/(z-z_0)$\cite{2}. Since the chiral anomaly is not corrected in the heterotic case, the number of zero modes should be still four. We now proceed to derive the Dirac equations for $\psi$ and $\zeta$ from the heterotic Lagrangian and Wick-rotate them to Euclidean space by $x^0\rightarrow -ix^2$.  We only keep linear terms in the equations of motion and solve them. We will see that the solutions are also satisfied with the equations of motion when we add nonlinear terms. Since we have already shown that the number of independent zero modes is four, the solutions solved in the equations of motions up to linear terms are all zero modes. From Eq.\ (4) and instanton background $\bar{\partial}\phi=0$, in spinor components we have:
\begin{eqnarray}
\delta\zeta^{\dagger}_R&\Longrightarrow&\bar{\partial}\zeta_{R}=0\\
\delta\zeta_{R}&\Longrightarrow&\bar{\partial}\zeta^{\dagger}_R+2\gamma\frac{\bar{\partial}\phi^{\dagger}}{\chi^2}\psi_{R}=0\\
\delta\psi^{\dagger}_{L}&\Longrightarrow&\partial\psi_{L}-2\frac{\phi^{\dagger}\partial\phi}{\chi}\psi_{L}=0\\
\delta\psi^{\dagger}_{R}&\Longrightarrow&\bar{\partial}\psi_{R}=0\\
\delta\psi_{R}&\Longrightarrow&\gamma g_{0}^{2}\zeta_{R}\bar{\partial}\phi^{\dagger}-\bar{\partial}\psi^{\dagger}_R+\frac{2\phi\bar{\partial}\phi^{\dagger}}{\chi}\psi^{\dagger}_{R}=0\\
\delta\psi_{L}&\Longrightarrow&\partial\psi^{\dagger}_{L}=0
\end{eqnarray}
Notice that Eqs.\ (8) (9) and (11) are the same as in the standard CP(1) case. Therefore the solutions to these three equations should be the same as before. They give all four zero modes. Since we have argued that the total number of zero modes is four, there is no extra zero modes raised from field $\psi^{\dagger}_R$, although the equation of motion of $\psi^{\dagger}_R$\ (10) is deformed by the $\gamma$ term. Therefore we have: 
\begin{eqnarray}
&&\psi_{R}^{(1)}=\frac{\rho}{(z-z_{0})^{2}}\,, \ \ \ \ \psi_{L}^{(1)}=0\,, \ \ \ \ \psi^{\dagger(1)}_R=0\,, \ \ \ \ \psi^{\dagger(1)}_L=0\,,\\
&&\psi_{R}^{(2)}=\frac{\rho}{(z-z_{0})}\,, \ \ \ \ \ \ \psi_{L}^{(2)}=0\,, \ \ \ \ \psi^{\dagger(2)}_R=0\,, \ \ \ \ \psi^{\dagger(2)}_L=0\,,\\
&&\psi_{R}^{(3)}=0\,, \ \ \ \ \psi_{L}^{(3)}=0\,, \ \ \ \ \psi^{\dagger(3)}_R=0\,, \ \ \ \ \psi^{\dagger(3)}_L=\frac{\bar{\rho}}{(\bar{z}-\bar{z_0})^2}\,,\\
&&\psi_{R}^{(4)}=0\,, \ \ \ \ \psi_{L}^{(4)}=0\,, \ \ \ \ \psi^{\dagger(4)}_R=0\,, \ \ \ \ \psi^{\dagger(4)}_L=\frac{\bar{\rho}}{(\bar{z}-\bar{z_0})^2}\,,
\end{eqnarray}
By the same argument, fermionic field $\zeta$ and $\zeta^{\dagger}$ cannot provide extra zero modes. From Eqs.\ (6) (7) and (10), we get:
\begin{eqnarray}
\zeta^{(1,2,3,4)}_{R}=0\,, \ \ \ \ \zeta^{\dagger(1,2,3,4)}_R=-2\gamma\,\frac{\phi^{\dagger}}{\chi}\,\psi_R^{(1,2,3,4)}
\end{eqnarray}
At first sight, the solutions seem inconsistent with the anomaly calculation, which gives the number of the difference between left and right handed fermions under instanton background, $i.e.$
\begin{eqnarray}
&&\int d^{2}x\partial_{\mu}j^{\mu
}_5=\int d^{2}x\frac{2i}{\pi}\epsilon^{\mu\nu}\frac{\partial_{\mu}\phi^{\dagger}\partial_{\nu}\phi}{\chi^2}=4
\end{eqnarray}
From solutions (12)-(15), It seems that the first two solutions are right handed fermions while the last two are left handed, thus the number of the difference is zero. However, because we wick-rotated to Euclidean space to perform the calculation, in which the complex conjugate is not a well-defined involution for chiral fermions, the field $\bar{\psi}$ is thus totally independent of $\psi$. Moreover in Euclidean space, the Lagrangian is $SO(2)$ invariant. Therefore the field $\bar{\psi}$ transformed under $SO(2)$ is the same as $\psi^{\dagger}$. Bearing this in mind, we can be back to the Lagrangian (4):
\begin{eqnarray} 
\mathcal {L}_{\psi}&=&\frac{2i}{g_{0}^2\chi^2}\bar{\psi}\gamma^{\mu}(\partial_{\mu}\psi-\frac{2}{\chi}\phi^{\dagger}\partial_{\mu}\phi\psi)\nonumber\\
&=&\frac{2i}{g_{0}^2\chi^2}\bar{\psi}_L\gamma^{\mu}(\partial_{\mu}\psi_R-\frac{2}{\chi}\phi^{\dagger}\partial_{\mu}\phi\psi_R)+\frac{2i}{g_{0}^2\chi^2}\bar{\psi}_R\gamma^{\mu}(\partial_{\mu}\psi_L-\frac{2}{\chi}\phi^{\dagger}\partial_{\mu}\phi\psi_L)
\end{eqnarray}
Now it is clear that, when wick-rotated to Euclidean space, the field $\psi_R^\dagger$ and $\psi_L^\dagger$ we solved in Eqs.\ (10) and (11) are in fact the fields $\bar{\psi}_L$ and $\bar{\psi}_R$ respectively, while the field $\bar{\psi}$ plays the same role as $\psi^{\dagger}$ in Euclidean space. Therefore the four solutions are all right handed fermions. It is consistent with the anomaly calculation.

As for the field $\zeta$, solution (16) shows that the extra field $\zeta$ mixes with field $\psi$, however it is linearly dependent of field $\psi$, and thus does not produce extra zero modes as what we also expected from the result of the chiral anomaly calculation (17).

At last, one more worthy to mention is the solutions of field $\psi$. They are of the same form as in the standard {CP(1)} zero modes. It means that not only the number of zero modes does not change, the zero modes of $\psi$ also receive no correction in the heterotic model. In fact it is easy to understand this result from equation (8) and (9). Although the supersymmetries are partly broken by deformation in the heterotic case, the classical equation of motion of $\psi$ up to linear terms are still $\mathcal{N}=(2,2)$ supersymmetic invariant. Therefore in this sense the zero modes of $\psi$ still can be generated by supercharges and superconformal charges from the instanton background, even though half of the symmetries are broken in the heterotic model. In addition, one can check that these solutions are satisfied with the equations of motion when adding nonlinear terms of fermions.\\

\section{Heterotic CP(N-1) Sigma Model:}

Our conclusion to the heterotic {CP(1)} model is that the zero modes of $\psi$ are the same as in the standard {CP(1)} case, while the extra fermion $\zeta$ does not provide extra zero modes. This result can be simply generalized to the heterotic {CP(N-1)} model. For the heterotic {CP(N-1)} case, the Lagrangian containing bilinear terms of fermions is\cite{1}:
\begin{eqnarray}
\mathcal
{L}&=&\zeta_{R}^{\dagger}i\partial_{L}\zeta_{R}+[\gamma
g_{0}^{2}\zeta_{R}G_{i\bar{j}}(i\partial_{L}\phi^{j\dagger})\psi_{R}^{i}+h.c.]
+iG_{i\bar{j}}\bar{\psi}^{\bar{j}}\gamma^{\mu}(\partial_{\mu}\psi^{i}+\Gamma^{i}_{lk}\partial_{\mu}\phi^{l}\psi^{k})
\end{eqnarray}
Hence, the equations of motion up to linear terms under instanton backgroud $\bar{\partial}\phi^{i}=0$ are:
\begin{eqnarray}
\delta\zeta^{\dagger}_{R}&\Longrightarrow&\bar{\partial}\zeta_{R}=0\\
\delta\zeta_{R}&\Longrightarrow&\bar{\partial}\zeta^{\dagger}_R+\gamma g_{0}^{2}G_{i\bar{j}}\bar{\partial}\phi^{j\dagger}\psi_{R}^{i}=0\\
\delta\psi_{L}^{\dagger\bar{j}}&\Longrightarrow&\partial\psi_{L}^{i}+\Gamma^{i}_{kl}\partial\phi^{l}\psi_{L}^{k}=0\\
\delta\psi_{R}^{\dagger\bar{j}}&\Longrightarrow&\bar{\partial}\psi_{R}^{i}=0\\
\delta\psi_{L}^{j}&\Longrightarrow&\gamma g_{0}^{2}\zeta_{R}\bar{\partial}\phi^{\dagger\bar{i}}-\bar{\partial}\psi^{\dagger\bar{i}}_R-\Gamma^{\bar{i}}_{\bar{k}\bar{l}}\bar{\partial}\phi^{\dagger\bar{l}}\psi_{L}^{\dagger\bar{k}}=0\\
\delta\psi_{R}^{j}&\Longrightarrow&\partial\psi_{L}^{\dagger\bar{i}}=0
\end{eqnarray}
It is not hard to see the solutions are:
\begin{eqnarray}
&&\psi_{R}^{i(1)}=\frac{\rho^{i}}{(z-z^{i}_{0})^{2}}\,, \ \ \ \ \psi_{L}^{i(1)}=0\,, \ \ \ \ \psi^{\dagger\bar{i}(1)}_R=0\,, \ \ \ \ \psi^{\dagger\bar{i}(1)}_L=0\,,\\
&&\psi_{R}^{i(2)}=\frac{\rho^{i}}{(z-z^{i}_{0})}\,, \ \ \ \ \ \ \psi_{L}^{i(2)}=0\,, \ \ \ \ \psi^{\dagger\bar{i}(2)}_R=0\,, \ \ \ \ \psi^{\dagger\bar{i}(2)}_L=0\,,\\
&&\psi_{R}^{i(3)}=0\,, \ \ \ \ \psi_{L}^{i(3)}=0\,, \ \ \ \ \psi^{\dagger\bar{i}(3)}_R=0\,, \ \ \ \ \psi^{\dagger\bar{i}(3)}_L=\frac{\bar{\rho^{i}}}{(\bar{z}-\bar{z_0^{i}})^2}\,,\\
&&\psi_{R}^{i(4)}=0\,, \ \ \ \ \psi_{L}^{i(4)}=0\,, \ \ \ \ \psi^{\dagger\bar{i}(4)}_R=0\,, \ \ \ \ \psi^{\dagger\bar{i}(4)}_L=\frac{\bar{\rho^{i}}}{(\bar{z}-\bar{z_0^{i}})^2}\,,\\
&&\zeta_R=0\,,\\
&&\zeta_R^{\dagger(1,2,3,4)}=-\gamma g_0^2\,\frac{\partial K}{\partial\phi^{i}}\,\psi_{R}^{{i}(1,2,3,4)}
\end{eqnarray}
where $K$ is the K\"{a}hler potential. From these solutions we see that the number of zero modes in the heterotic {CP(N-1)} model is still $4(N-1)$, while $\zeta$ can be linearly expressed by the zero modes of $\psi$ and thus not extra zero mode.\\

\section*{Acknowledgements}
I am grateful for M. Shifman who led me to the subject, and A. Vainshtein for very useful and fruitful discussions and reviewing the final version of the manuscript. 
I also want to thank M. Sasseville for proofreading the text.\\


\begin{thebibliography}{99}

\bibitem{4} B. Zumino, Phys. Lett. {\bf B87}, 203 (1979)
\bibitem{5} M. Edalati and D. Tong, JHEP {\bf 0705}, 005 (2007)
\bibitem{1} M. Shifman and A. Yung, Phys. Rev. {\bf D77}, 125016 (2008)
\bibitem{2} V. A. Novikov, M. A. Shifman, A. I. Vainshtein and V. I. Zakharov, Phys. Rept. {\bf 116}, 103 (1984)
\bibitem{3} M. Shifman, A. Vainshtein and R. Zwicky, J. Phys. {\bf A39}, 13005 (2006)



\end{thebibliography}
\end{document}